# The role of the US National Office in the Gemini partnership

Kenneth H. Hinkle*[a], Letizia Stanghellini[a], Dara Norman[a], Sharon Hunt[a]

[a]National Optical Astronomy Observatory, 950 N. Cherry Street, Tucson, AZ USA 85719-4933

**ABSTRACT**

We follow the history of the US National Gemini Office from its origin when the US National New Technology Telescope was reshaped into two 8m telescopes for the International Gemini Observatory. The development of the office in the decade of the 1990s continues to shape its function to the present. The following decade, 2000–2010, marked major milestones including the dedication of the telescopes, the reshaping of the Gemini instrumentation program, and dissatisfaction of the US community as expressed in the ALTAIR report. Nationally funded facilities are under financial pressure, as new projects must be funded from a nearly fixed budget. We will discuss how the US NGO should be used to advocate for both the US community and the Gemini Observatory. This role could be an essential one in protecting open access to 8m-class facilities.

## 1. INTRODUCTION

Since the 1950s, the US National Science Foundation (NSF) has funded telescopes to provide access to the entire astronomical community to major facilities through a peer review process. The Association of Universities for Research in Astronomy (AURA) was founded in 1957 to operate a NSF-funded national astronomical observatory. This effort led to the formation of Kitt Peak National Observatory (KPNO) and shortly thereafter Cerro Tololo International Observatory (CTIO). These two observatories were funded separately by the NSF until they were combined into the National Optical Astronomy Observatory (NOAO) in 1983.

In the early 1980s, NOAO started planning for a National New Technology Telescope (NNTT) that would far exceed the light-gathering power of the existing 4m telescopes. While this was originally a 15m telescope project, technical and fiscal pressures resulted in a 1987 recommendation by an AURA committee to build two 8m-class telescopes, one in each hemisphere. The US Congress passed legislation in September 1990 to fund an international collaboration. US funding was capped at 50% or $88 million, whichever was the least [1]. The outcome of this process was the Gemini Observatory with US, UK, and Canadian partners.

Twenty-eight years later, the NSF approved a major reorganization of AURA-managed ground-based observatories. With the close of FY18, NOAO will cease to exist and will be reorganized with Gemini and LSST Operations into the

---

*hinkle@noao.edu; phone 1 520 318-8298; noao.edu

National Center for Optical-Infrared Astronomy (NCOA). While Gemini will remain a branded entity with an international partnership, it will share a management structure and staff with other NCOA units.

One feature of the international Gemini partnership has been that each partner country maintains a National Gemini Office (NGO) to advocate for the telescope in that country and to provide the first level of user support. What is the role of the US NGO in a corporate structure where the US NGO and Gemini are both operated by the same organization? What is its future in an organization that will likely be focused on the LSST mission? To gain insight we start by reviewing the history of the US NGO. This is summarized on a timeline in Figure 1.

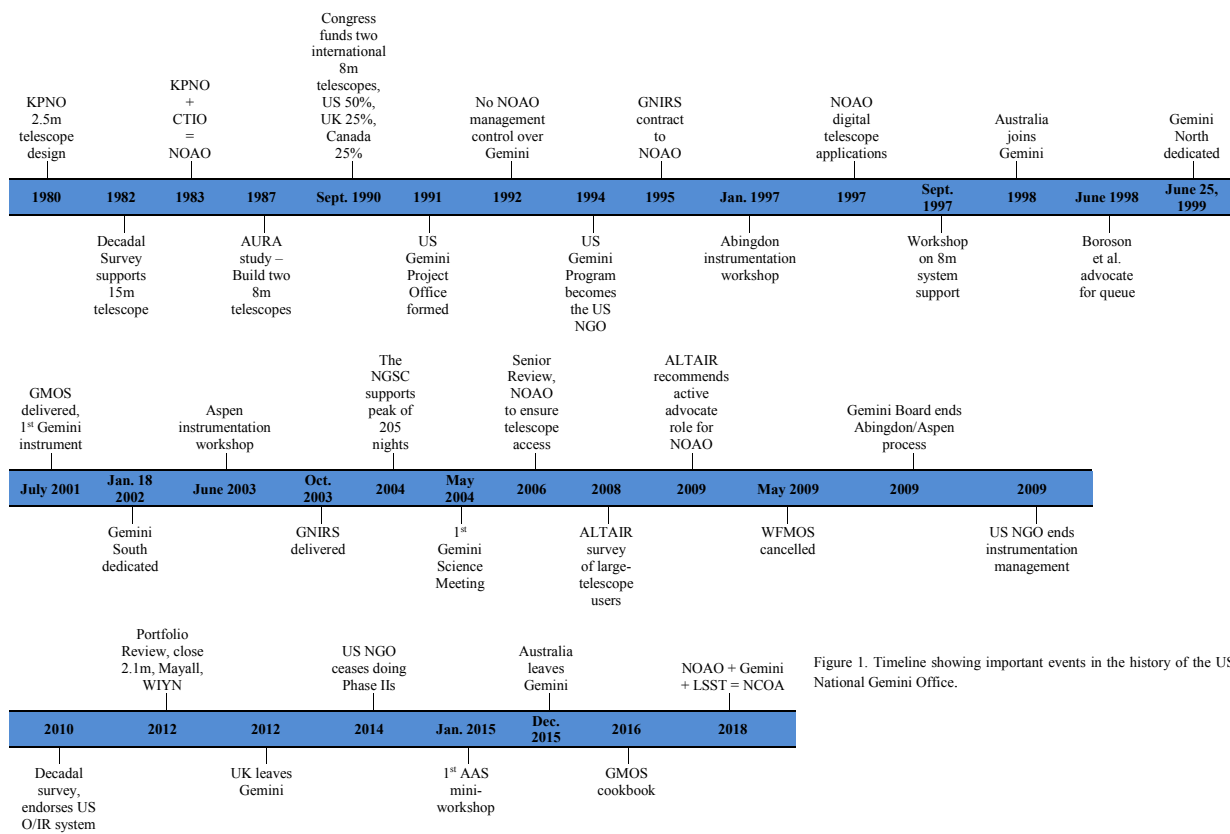

Figure 1. Timeline showing important events in the history of the US National Gemini Office.

## 2.  US GEMINI PROJECT OFFICE

Until 1990, the 8m telescope project was a US program operated by NOAO. By 1992, the Gemini Board "formally stopped NOAO management exercising any direct control over the project" [1]. NNTT and NOAO were faced with a difficult transition period. A US project office was formed in NOAO-Tucson in 1991 to support the participation of US astronomers, including NOAO staff, in Gemini advisory committees. A Science Advisory Committee (SAC) along with instrument working groups were identified in that year's annual report. A formal arrangement followed in the next two years. Each of the Gemini partner countries was to set up a National Project Office "to act as a focus for national scientific and technical interests within their countries" [2]. These offices were to be regarded as an integral part of the Gemini Project. At the annual 1993 meeting of the AURA Board, it was resolved to establish a US Gemini Program Office (USGPO) as a separate division within NOAO on par with the ground-based 4m (KPNO and CTIO) observatory offices [2].

The 1993 NOAO Annual Report contains a lengthy discussion of the activities of the USGPO. There was considerable overlap of activities taking place at NOAO and Gemini. Fred Gillett was the acting US Gemini Project scientist, supported by Jay Gallagher and Richard Green from the NOAO staff, an administrative assistant, and an engineer/manager. Gillett played a key role in advocating for infrared optimization of the Gemini telescopes and personally carried out many of the detailed engineering calculations [3]. Within a year Gillett left NOAO to take a position with Gemini.

The 1993 NOAO Annual Report highlights the following activities of the USGPO. Many of these are technical or engineering in nature.

- Community Input. Through a connection to the US SAC, community input was sought on the telescope specifications. The USGPO solicited participation by the US astronomy community in building instruments for Gemini.

- Project Involvement. The USGPO together with the Gemini project worked to characterize the scientific performance of the telescope and to solve technical problems. The USGPO coordinated external review of technical reports.

- Dissemination of Information. Articles were submitted to both the Gemini and NOAO newsletters. The USGPO had displays at the AAS and ASP meetings. The USGPO staff started a series of presentations on scientific and technical aspects of Gemini. An archive was started for Gemini Technical Reports.

- Technical Activities. The USGPO commissioned a model telescope and enclosure and oversaw the testing of optics.

This level of activity continued into 1994, with the NOAO Annual Report listing USGPO participation in numerous CDR, CoDR, PDR, working group, and development meetings.

## 3. US GEMINI PROGRAM

The US National Gemini Office that exists in 2018 was born in 1994. That year, the USGPO was renamed the US Gemini Program (USGP) with the role of the USGP focused on being a liaison between the International Gemini Observatory (IGO) Project Office and the US community. By 1994, the USGP scientific staff were engaged in giving colloquia on the 8m telescope project around the US and had a booth at major meetings. This year also saw the start of an email newsletter that was sent every three weeks providing updates to the quarterly *NOAO Newsletter*. With the departure of Fred Gillett to the IGO staff, Todd Boroson became the USGP scientist. Boroson's tenure with the USGP lasted until 2000, spanning the transition of the Gemini Observatory from early construction funding to an observatory with functioning telescopes.

## 3.1 Governance

A point that frequently leads to confusion is that the USGP is not part of Gemini. The connection of the US NGO to Gemini takes place through the head of the NGO, who has membership on the Gemini Operations Working Group (OpsWG). The OpsWG advises the Gemini director. This indirect connection results largely from the governance structure of Gemini. The US NGO is part of NOAO. NOAO is funded by the NSF through a grant to AURA. AURA provides review of NOAO through a board that appoints visiting committees and financial review committees.

Gemini, while operated by AURA, is governed differently. As set out in the Gemini International Agreement (www.gemini.edu/science), authority for operating Gemini rests with the Gemini Board. The number of board representatives reflects the partnership fractions. These fractions have changed over the years. Board membership in 2018 consists of six representatives from the US, two from Canada, and one each from Hawaii, Chile, Argentina, and Brazil. US board members are selected by the NSF. Typically, the NSF does not solicit input from NOAO when selecting board members. Scientific and technical input to the board comes from the Gemini Science and Technology Advisory Committee. The Gemini Finance Committee advises the board on financial, budget, and planning issues. Partner funding for Gemini flows though the NSF to AURA, Inc. An AURA Oversight Council provides management oversight.

## 3.2 Instrumentation

One departure from the liaison role for the USGP was the acquisition of new instrumentation. The IGO agreement with the National Gemini Offices called for the first-generation instruments to be "allocated separately, in proportion of the contribution of the Partners" [4]. The US was to provide three instruments, a near-IR (1–5 microns) imager, a near-IR spectrograph, and a mid-IR (8–30 microns) imager. The USGP had oversight responsibilities for the instruments, with shared responsibilities with the IGO for reviews and acceptance tests. The USGP was also responsible for procurement of all the IR arrays, IR controllers, and CCDs.

The NSF assigned the near-IR imager (NIRI) to the University of Hawaii. The USGP solicited proposals for the Near-IR spectrograph (NIRS) in 1994. Construction of this instrument was awarded to NOAO in 1995 [5]. The contract for the mid-IR imager was awarded in 1997. For this instrument, the USGP carried out a two-stage procurement process. Funded conceptual design studies were incorporated into an RFP for the final design and construction of the instrument. While the USGP ran the process, it did not take part in the evaluation of or selection of proposals.

At the same time the first-generation instruments were being procured, planning for the second-generation instruments started. Several workshops on adaptive optics were organized by the USGP from 1995 through 2001. In 1996, the USGP held a workshop on future instruments for the Gemini telescopes. This was followed in January 1997 by the IGO workshop on instrumentation in Abingdon, England. USGP supported participation in this meeting by US astronomers. At the Abingdon meeting, the discussion focused on science goals in order to form a consensus on future directions for Gemini instrumentation [6].

The Abingdon meeting was followed by a considerable increase in instrument planning and procurement. In 1998, the Committee of Gemini Offices (CGO), comprised of NGO scientists, NGO project managers, and IGO staff, set up an instrument forum as part of the semi-annual meetings. The instrument forum was used to formulate instrument plans and to make recommendations for groups to design and build instruments. The CGO depended on proposals solicited by the NGOs. In March 1998, this process resulted in evaluation of proposals for a polarization module and an AO system. In September 1998, the CGO discussed allocating conceptual design studies for a near-IR coronagraph/imager. Five US groups were interested and USGP held a review of concepts. One group was funded for a design study. Also discussed was a second near-IR spectrograph. This resulted in a study of micro-mirrors by NOAO, STScI, and Goddard Space Flight Center.

In 1999, the near-infrared spectrograph NIRS encountered significant problems in design and management resulting in large cost overruns and delays. A joint review conducted by IGO and NOAO was hosted by AURA. One outcome was a review of USGP management of instrumentation procurements. US Gemini instruments were managed directly through the USGP, and a number of programs were in progress. In 2000, there were four Gemini projects at NOAO—GNIRS, Gemini IR array controllers, GMOS/HROS CCDs, and Phoenix to Gemini upgrade—and two external US contracts— a near-IR coronagraphic imager (NICI) and T-ReCS, a mid-IR imager and spectrograph with the University of Florida. NICI was funded with NASA money to NOAO that was then used to fund building NICI at Mauna Kea Infrared (MKIR). In 2000, a contract was being negotiated with the University of Florida to build a near-IR imager and multi-object spectrograph called FLAMINGOS 2 (F2).

In 2001, the IGO decided to start taking a direct role in the procurement of future instruments. By 2002, the role of the USGP in instrumentation was being gradually reduced. Eventually the role was limited to contract oversight, and in a decade, it was entirely eliminated.

### 3.3 Operations

The ramp-up of Gemini facilities and infrastructure toward operation started in earnest in 1997 [7] with the operations hand-over of Gemini North scheduled for June 2000. The USGP initiated planning for a US National Gemini Telescope Allocation Committee (TAC), a US National Gemini User Committee, proposal writing assistance, Phase II assistance, and data reduction assistance. USGP also started long-range planning for a remote observing center. In 1997, at the behest of the USGP, the NOAO telescope time application form, which had been a paper form, was converted to a web-based or emailed LaTeX form.

In 1998, the CGO held an operations forum to explore issues associated with Gemini operations. The timing of the first call for proposals, a data archive, outreach, and support for observing-related information requests were discussed. An electronic help desk for providing user support was discussed. The USGP (and its encompassing NOAO division, Science Operations Division or SCOPE) believed they had responsibility for seven items:

- Providing information to the US astronomy community
- Submission, processing, and review of telescope time proposals
- Assistance in proposal preparation, pre-observing support, and post-run support including data reduction
- Providing reduction software
- Developing a data archive
- Investigating modes of operation, i.e., classical, remote observing, queue, etc.
- Outreach and dialogue with the community

Boroson was concerned that with the focus on 8m-class telescopes smaller-scale facilities would be abandoned. The USGP organized a three-day workshop in Tucson in September 1997 to identify and quantify the supporting facilities needed for the US community to use Gemini telescopes effectively. The premise was that telescopes are systems where the largest-telescopes proposals are based on data, especially surveys, obtained by smaller telescopes. The meeting goal was to quantify these ideas using a science-based approach. Fifty astronomers, mainly from the US, worked in panels devising major programs for 8m-class telescopes. They then analyzed these programs to understand all the necessary capabilities other than the 8m telescopes. The process lead to unanimous recognition that wide-field surveys, 4m-class observations, detector development, software pipelines, and accessible archives were key needs [8].

In 1998, a USGP workshop was held on the proposal process that included a number of public observatories. Merging processes for combining grades submitted by separate TAC committees were under development. A working group reported on survey programs for large telescopes.

The Gemini North telescope was dedicated on 25 June 1999, and Gemini South was dedicated on 18 January 2002 [1]. The Gemini North telescope started science operations in June 2000. The initial set of proposals for Gemini North from the US community was for shared-risk and early science with a factor of six oversubscription. To celebrate the first data from Gemini, the USGP and the Canadian Gemini project office co-hosted a special session at the June 2001 AAS meeting on "First Science from the Gemini Telescopes." The first Gemini facility instrument, GMOS, was delivered in July 2001.

In 2001–2002, the USGP had twelve scientific staff, albeit mostly shared with other NOAO divisions, and four technical or administrative staff who covered a range of diverse activities involving the USGP. USGP staff were involved in engineering runs at the telescopes, demonstration science, and observing support. USGP continued to run a US Gemini Science Advisory Committee. A subset of this committee participated in Gemini Science Advisory Committee meetings. The head of the USGP attended Gemini Operations Working Group (OpsWG) meetings. The USGP was assigned primary responsibility for providing data reduction software. The software was largely drawn from the existing IRAF software package with extensions as needed. IGO was developing a web-based help desk [9]. The USGP was to be the first tier to answer US questions.

### 3.4 The queue

Since near the inception of the Gemini project, operational plans foresaw a mix of remote and queue observing [4]. An international workshop on observing modes was held in 1995. To test the queue mode for a ground-based telescope, USGP and WIYN staff carried out queue-scheduled operations at the Kitt Peak WIYN telescope. WIYN operations can be similar to those at Gemini since multiple instruments can be used on any night. After two years of queue operation, Boroson et al. [10] reported on a comparison of queue and classical scheduling. They found that the queue did deliver the improvements predicted in simulations. The overall telescope efficiency is about 15% higher than it would have been if classically scheduled. Roughly four times as many observations requiring the best seeing were obtained by the queue as would have been obtained otherwise. The queue resulted in approximately 2.5 times more programs being completed as in the classical comparison. In spite of this result, questionnaires sent to astronomers applying for WIYN time found a preference for classical observing. Boroson and Trueblood [11] wrote: "The queue approach tends to polarize users' views: those who received data were pleased with its quality and with the experience, while those who were in the queue but did not receive data would rather have come to the telescope and taken their chances. By stressing the completion of programs, the queue results in a smaller number of users who get any data, and so the questionnaire responses seem to indicate overall dissatisfaction. It will be important for Gemini and the National Gemini Offices to understand how to address this sort of reaction."

## 4. NOAO GEMINI SCIENCE CENTER

In 2003, the USGP, renamed the NOAO Gemini Science Center (NGSC), was fully engaged in supporting the US Gemini community. This included answering HelpDesk questions by proposers and users, conducting technical reviews of proposals, supporting the NOAO TAC process, supporting the Phase II process, and providing selected operational support to Gemini. The NGSC supported operations of the NOAO Phoenix spectrograph at Gemini South and provided queue-observing support for GMOS-North and GMOS-South. Taft Armandroff, the director of the NGSC from 2002 through 2006, aggressively pushed the NGSC agenda on all fronts. The NGSC had booths at the AAS meeting to attract potential users and on-site user assistance. Webcasts were done before proposal deadlines to update the community on instrument availability. There were frequent teleconferences and email discussions with the US SAC in addition to yearly meetings. The advice of the US SAC was sought on the current state of observing capabilities on Gemini, on future opportunities, and on how the priorities of the US Gemini community should be enunciated.

The NGSC was also active in planning for future Gemini instruments. The NGSC hosted several meetings in early 2003, including one for the US SAC and a workshop on "Future Instrumentation for the Gemini 8-m Telescopes: US Perspective in 2003" that explored the science questions expected in the 2008–2010 period and the instrumentation needed to address these questions. This was in preparation for an IGO science and instrument planning meeting that was held in Aspen, Colorado, in June 2003. This was intended as the next major Gemini instrument planning meeting in the tradition set by the Abingdon meeting. Twenty-eight US delegates attended. The NGSC held briefings and discussions for the US delegates.

However, the role of the NGSC in the acquisition of Gemini instrumentation by this time was managerial. As a final step in the instrument contracts signed in the 1990s, NGSC representatives were present for final acceptance tests of US instruments. T-ReCS passed its final acceptance test in FY03. GNIRS was delivered to Cerro Pachón on October 31, 2003, and formally accepted by Gemini in February 2004. Since GNIRS was an NOAO project, the NGSC advocated for GNIRS by sponsoring a 2004 pilot program, "GNIRS Key Science Opportunity," for longer programs. NICI passed acceptance testing in FY07 and became a Gemini facility instrument in 2009A. F2 underwent acceptance testing and had first light in 2009 (https://www.gemini.edu/node/11328). This ended direct involvement by the US National Gemini Office in the Gemini instrumentation program.

The Gemini Data Reduction Working Group (DRWG) was founded in 2006 with a NGSC staff member as chair. They met and produced a report that same year. DRWG comments were requested on the Gemini pipeline project in 2009; they were not supportive of this effort. The DRWG has not convened since. However, in 2010 the first of a series of Gemini Data reduction workshops was held.

### 4.1 Hands-on observing support

A major and growing responsibility of the NGSC in the early 2000s was support of users preparing Phase II plans. Gemini Phase II plans are detailed sets of instructions that executed the observations in the queue. If a program was scheduled, regardless of band, the PI was required to submit a Phase II plan or the proposal would be "de-queued." Since the Gemini telescopes were largely queue scheduled, targets specified in a user's observing program are not observed in blocks. Many programs have targets that can be observed throughout the semester and they typically were. As a result, Gemini would set a deadline for all observers to submit their Phase II plans weeks before the start of the semester and a deadline for these Phase II plans to be activated in the queue by the start of the semester. The time between the deadlines was typically a few weeks, and the total time between PI notification and activation was typically less than two months. Due to the typical human response of procrastination, the NGSC needed to have enough staff to check all the programs and interact with observers in two to three weeks. This resulted in a large scientific staff involvement for a few weeks twice a year. Many of the staff had no other interaction with Gemini for the rest of the semester.

The NGSC was also providing observing support at Gemini telescopes. Plans for remote operations never occurred because of concerns that the telescope operating system could be hacked. In 2001, USGP staff carried out mini-queue observations with visiting instruments, OSCIR and Hokupa'a/QUIRC, at Gemini North. Starting in 2002, USGP staff were supporting Phoenix operations on Gemini South. This initially involved five staff. At that time Phoenix was the most requested instrument at Gemini South. USGP staff also provided some queue-observing support with NIRI and in the Hokupa'a mini-queue runs. In FY04, NGSC staff were at Gemini telescopes for 205 nights either assisting with observing or doing engineering tests. In 2005, NGSC staff covered 174 nights and in FY06 they covered 111 nights. In FY07, Phoenix was supported by NGSC staff on 27 nights, and NGSC staff helped with the queue for other instruments on 64 nights.

The Gemini telescopes are designed to have four instruments mounted and ready for observations, with queue observations from any one of these possible. This presented an obstacle for NGSC observing support. The queue observer needed to be checked out and familiar with observing with each instrument in the queue. As more instruments were commissioned, this was not practical for NGSC staff who visited the Gemini telescopes one or two times a semester, and hands-on NGSC support ceased.

The first Gemini Science Meeting was in Vancouver, British Columbia, in May 2004. The Gemini Science Meetings are intended to discuss current results, capabilities, and future research and collaborations related specifically to Gemini. Gemini decided to hold this series of meetings every three years, with the next meeting in Foz de Iguaçu, Brazil, in June 2007. The partner National Gemini Offices started meeting in 2004, with meetings every 18 months. The third meeting coincided with the Foz de Iguaçu Gemini Science Meeting.

As an outcome of the Aspen process in December 2003, Gemini issued a call for proposals for instrument studies for a high-resolution near-infrared spectrograph. The NGSC was not officially involved, but several of the NGSC staff did submit a proposal. A joint proposal from NOAO/UF ultimately won the competition, but Gemini then withdrew funding. NOAO made a large investment of engineering staff in preparing the proposal. In 2008, Gemini and Subaru continued to discuss the Wide-Field Multi-Object Spectrograph (WFMOS). Conceptual design studies were underway by two teams. While the NGSC was not directly involved, NOAO staff were involved, and the NGSC did fund travel to a meeting in May 2008 on "Cosmology Near and Far: Science with WFMOS." The third Gemini Science Meeting was held jointly with Subaru in Kyoto, Japan, in May 2009. This meeting was expected to launch the WFMOS project, but instead the Gemini director announced that Gemini would not go ahead with WFMOS due to funding problems. There had been a considerable investment of time and resources by various US groups, by Gemini, and by international collaborators. At approximately the same time, the Gemini Board terminated the Aspen instrument program. There had been ongoing studies of high-resolution bench-mounted near-IR spectrographs by university groups. These were stopped. The one instrument that came out of the Aspen process was the Gemini Planet Imager (GPI).

### 4.2 ALTAIR

The NSF 2006 Senior Review Committee (https://www.nsf.gov/mps/ast/seniorreview/sr-report.pdf) encouraged NOAO to ensure scientifically balanced community access over all telescope apertures. Todd Boroson, US Gemini Program director in the 1990s, was the NOAO director in 2007–2008. As part of the NOAO response, Boroson organized a community-based committee, "Access to Large Telescopes for Astronomical Instruction and Research" (ALTAIR), charged with studying the capabilities needed on 6.5m to 10m telescopes in 2010–2020 [12]. Verne Smith, who in 2006 had taken over as head of the NGSC, was the NOAO liaison. One other NGSC staff member was on the committee. A large part of the committee's focus was on understanding both the current and future role of the Gemini telescopes in the US ground-based O/IR system [13].

ALTAIR surveyed US astronomers on current and future use of US 6.5m to 10m telescopes, including aspects such as required instrumentation, observing modes, and observing time [13]. The large number of responses received (over 500) showed the presence of a large, interested, and energized community of large-telescope users. *NOAO Currents* [14] reported, "One of the most striking results of the survey was the response to the essay-format question that asked, 'If you have used both Gemini and non-federal facilities, how does Gemini compare with these facilities?' Many respondents commented thoughtfully and passionately in response to this question. Overall, the responses revealed broad dissatisfaction with Gemini, with Gemini characterized as uncompetitive or lagging relative to other large-telescope facilities. Respondents remained fundamentally supportive of Gemini, but they found their experiences with Gemini frustrating and were disappointed that Gemini does not perform better." On the other hand there was support for increasing the US share of Gemini.

Gemini received good marks from the survey on delivered-image quality, infrared performance, queue scheduling, and the ability to accommodate targets of opportunity. There was strong criticism that Gemini instrumentation was not competitive with instrumentation available on other large telescopes and was not aligned with the needs of a broad community of users. Several respondents said that Gemini would be a higher priority for them if it were more responsive to the needs of the US community. The second most common concern, which also elicited passion, was the large amount of time that a proposer had to spend at all stages of the process to end up with data. Respondents commented on the time burden of the Phase II process, concerns about data quality and observing efficiency, and the possibility of receiving little useful or no data for the effort expended. Others expressed a strong desire to be more actively involved in collecting their own data through classical observing or a remote eavesdropping process [14].

ALTAIR recommended that the NSF increase US participation in Gemini but only if Gemini became more responsive to the US community and evolved instrumentation, operations modes, and other services to be more aligned with the needs of the US community. ALTAIR also recommended that NOAO develop and maintain a roadmap for the development of the large-telescope system based on regular input from the US community and that NOAO be an active advocate for the development of the large-telescope system, using tools such as input to the Gemini Board and time trades with other observatories to achieve a balance of publicly available capabilities [15].

A root cause for dissatisfaction with Gemini was identified as the very limited role that the US community had in setting scientific goals for Gemini. In addition to the roadmap, ALTAIR recommended that the NSF change the Gemini governance structure and representation on the Gemini Board and Gemini Science Committee and create pathways for community input to the board. In response, the NGSC convened several meetings of a US Gemini Caucus, comprised of US representatives on the Gemini Board, Gemini STAC, the chair of the NOAO Users Committee, the NSF program officers for Gemini and NOAO, the NOAO director and deputy director, the NOAO System Science Center director, and the head of NOAO SUS. The intent was to have a group focused on the US Gemini perspective and to lobby the Gemini Observatory (http:/ast.noao.edu/about/committees/us-gemini-caucus). The Gemini Caucus was active for about a year. Another ALTAIR recommendation was for an improved instrumentation suite at Gemini that was more responsive to the needs of the US community. In response, the US Gemini Caucus worked to create a new instrument start for Gemini on a short timescale. There was strong demand for spectroscopic instrumentation. A survey asked users to prioritize an optical echelle, IR echelle, and an X-shooter (R~10000, UV to K). The X-shooter was top priority followed by an IR echelle [16].

The ALTAIR survey also asked about preferred observing modes. Both queue and classical were popular and valued observing modes. However, in a resource-limited environment, queue scheduling was viewed to be less critical than improvements to instrumentation or an increased number of nights for observing. Many respondents said that they liked the queue for its convenience [14]. However, many Gemini users also wanted to be more directly involved in taking their Gemini data. In response the NGSC encouraged classical observing, offering to cover the cost of travel expenses associated with Gemini classical observing [17].

## 5. THE CURRENT US NGO

For FY10, starting October 1, 2009, the NOAO Gemini Science Center was restructured into the NOAO System Science Center (NSSC) System User Support (SUS) group. The NGSC head, Verne Smith, was the head of NSSC, and Knut Olsen was appointed head of the SUS group. Letizia Stanghellini replaced Olsen in 2012. SUS was designed to support access to all facilities with US public-access time, but by far the largest share was from Gemini. Initially, SUS continued to support Phase II preparation and became involved in teaching data reduction fundamentals and detailed data reduction techniques for specific Gemini instruments. As part of this effort, the SUS was involved in the Gemini Data Reduction Workshop held in Tucson in July 2010.

In late 2009, SUS carried out a survey of Gemini users' instrumentation needs and presented the results at a "US Gemini Town Hall" meeting at the January 2010 AAS. SUS also responded to the request for white papers on a proposed high-resolution optical spectrograph for Gemini. Some NSSC staff participated in preparing a joint NOAO/Brazil Labóratorio Nacional de Astofísica proposal for the Gemini Optical Spectrograph, but this proposal was not selected for construction. 2011 saw the formation of the Gemini Science and Technology Advisory Committee (STAC), with one staff member from SUS appointed to the committee.

In 2014, continuing staffing pressures resulted in Phase II preparation no longer being done by NOAO staff. This function reverted to Gemini. Without the "seasonal" Phase II staff, the SUS staff were approximately 3 FTE in 2014 and shrank to 1.5 FTE by 2017. In 2015, the SUS name reverted to the US National Gemini Office (US NGO) since support was only being carried out for Gemini programs. As part of an increased focus on "Phase 3", i.e., post-run support, in 2015 the US NGO started maintaining a web portal (http://ast.noao.edu/csdc/usngo) referencing all documented reduction procedures for all Gemini instruments.

The US NGO also undertook the construction of a data reduction cookbook for the Gemini workhorse optical spectrograph GMOS. GMOS is a complex instrument, with multiple spectroscopy modes including long-slit, multi-slit, and integral field as well as an imaging mode and nod-and-shuffle CCD control. The cookbook deals with the variety of resulting data sets, and its development was a major effort by the US NGO (http://ast.noao.edu/sites/default/files/GMOS_Cookbook/).

The SUS / US NGO sponsored or collaborated in a number of meetings of interest to the Gemini user community. NOAO in collaboration with SUS held a workshop in 2013 on "Spectroscopy in the Era of LSST." Gemini Science Meetings continued with SUS / US NGO members giving various invited talks. In 2015, the US NGO started a series of mini-workshops at the winter AAS meeting on various Gemini topics of interest to the US user community (http://ast.noao.edu/csdc/usngo/mini-workshops).

Through the more than a decade since the dedication of the Gemini telescopes, the US NGO continues to carry out a number of routine functions. The US NGO provides the first response to HelpDesk questions. Until 2015, technical reviews were done on all Gemini proposals, but this was then switched to an as-needed basis. This change resulted not just from fewer US NGO staff but also from the increasing maturity of the user base. The NOAO Phoenix spectrograph has been offered at Gemini South nearly constantly since 2001. It was offered again at Gemini South in 2016, this time supported by the US NGO as a Gemini visiting instrument. At various times the SUS staff provided input to Gemini staff on the Gemini Integration Time Calculator, the Phase I tool (PIT), and the Phase II tool. Bi-annual OpsWG meetings continue, as do Gemini Science and Technology Advisory Committee meetings. There are biweekly video meetings held with the Gemini staff and NGOs. The NOAO TAC continued to review US Gemini proposals. SUS members work with other partner members to merge highly rated proposals with other partner countries at the International Telescope Allocation Committee. Starting in 2015, Gemini instituted Large and Long Programs with the TAC based at NOAO. In 2016, plans were developed for hosting the high-level science products from the Large and Long programs. Information on US NGO activities is provided to the community through its web pages, *NOAO Currents*, the *NOAO Newsletter*, and a US NGO presence at the winter AAS meeting.

## 5.1 Basic functions of the US NGO

During the decade after the formation of the International Gemini Observatory (IGO), goals were set for each National Gemini Office. These were summarized as a list of seven items in 1998. The list has been modified by ongoing agreements, the latest of which was made between NOAO and Gemini in 2017. While the current 2017 agreement provides a list of 28 core responsibilities, these can be distilled to the seven items on the 1998 list.

- Providing information to the US astronomy community

  *The 2017 contract notes maintaining the NGO web portal. It also notes the need to keep the representatives to committees current.*

- Submission, processing, and review of telescope time proposals

  *NOAO continues to run the TAC, while the US NGO does technical reviews.*

- Assistance in proposal preparation, pre-observing support, and post-run support including data reduction

  *The US NGO assists users with questions about proposal preparation and Gemini facilities. The US NGO no longer supports Phase II preparation. A directory of data reduction tools is maintained on the NGO web portal, and additional material is prepared and published as needed.*

- Providing reduction software

  *This has been replaced with a requirement to support post-observing data reduction.*

- Investigating modes of operation, i.e., classical, remote observing, queue, etc.

  *This is largely complete, but there is a current requirement to participate in the testing of user tools and software.*

- Outreach and dialogue to the community

  *This is mainly carried out through workshops and other meetings sponsored by the US NGO.*

## 6. FUTURE DIRECTIONS

Since the decade of the 1980s, the NSF budget has not been able to keep up with the demand for resources. This has resulted in pressure to close smaller telescopes in order to keep larger facilities open. The scientific impact discussed by Boroson [8] has been minimized due to the availability of ground- and space-based wide-field surveys and a system of private/university telescopes. However, the aperture size of telescopes being removed from public access has increased significantly. Casualties from the 2012 NSF Portfolio Review [18] were the Kitt Peak 2.1m, WIYN 3.5m, and Mayall 4m telescopes. While these telescopes were productively repurposed, their removal from the system impacted over 700 nights per year of open-access time [19]. Funding pressures are not limited to the US. Due to funding issues, the UK withdrew from Gemini at the end of calendar year 2012. Similarly, Australia joined in 1998 but withdrew in 2015. As devastating as the US cuts were, a more important message may be the rapidity of changes in direction. Following the 2010 Astronomy and Astrophysics Decadal Survey [20], NOAO was looking forward to "many more years of scientific productivity" for its 4m-class facilities [21]. The Decadal Survey had endorsed the US O/IR System and the role of NOAO in this system. Two years later the 2012 Portfolio Review called for the withdrawal of NSF funding for the Mayall.

The NSF astronomy division budget must increase at about 2.5% per year to fund the current facilities and LSST when operations start in a few years (Figure 2). Furthermore, the Elmegreen Report [22] urged the NSF to invest in one or both of the GMT and TMT telescope projects for LSST follow-up. If the NSF budget does not increase, the shortfall will be in excess of $30 million. Recently the NSF approved the reorganization of NOAO, LSST, and Gemini, all operated by AURA, Inc., into a unified management structure. One of the goals is resource optimization for support of the NSF flagship LSST survey. The Gemini South telescope is expected to be a major resource in LSST follow-up observing. While there will be savings, they will not approach the projected shortfall.

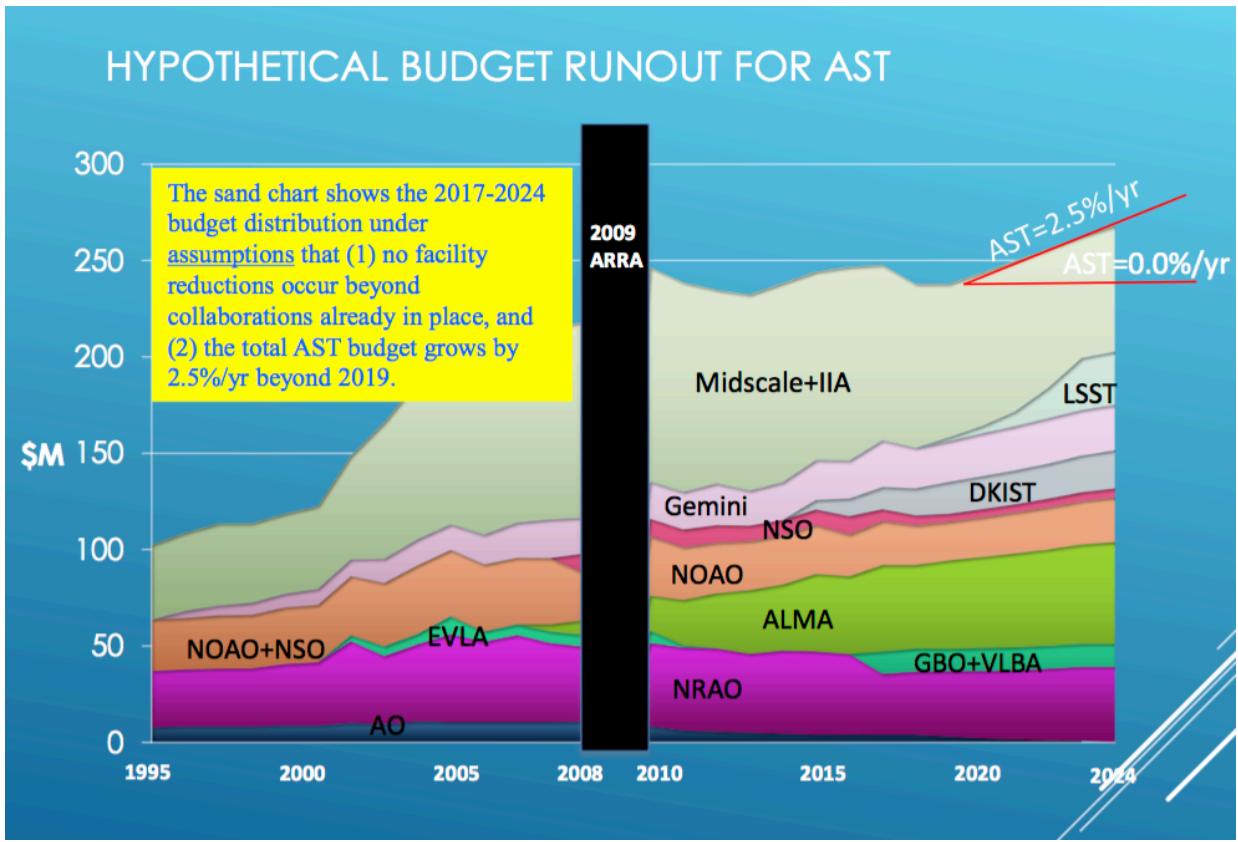

Figure 2. Hypothetical future budgets for the NSF MSP/AST division showing increased ALMA, DKIST, and LSST funding. The projections show a balanced budget if growth is 2.5% per year starting in 2020. Figure from Green [23].

Figure 3 shows the system of US telescopes in 2012. The Mayall and WIYN telescopes were major components. The figure shows the importance of both Gemini telescopes to the system. The Portfolio Review demonstrated that in times of tightening budgets even highly used nationally funded facilities could be in jeopardy. The US NGO was intended as the link between the IGO and the US community. Protecting the US open access nights on 8m-class telescopes should be a critical function. Here we summarize lessons from the past mission of the US NGO and discuss options for the future.

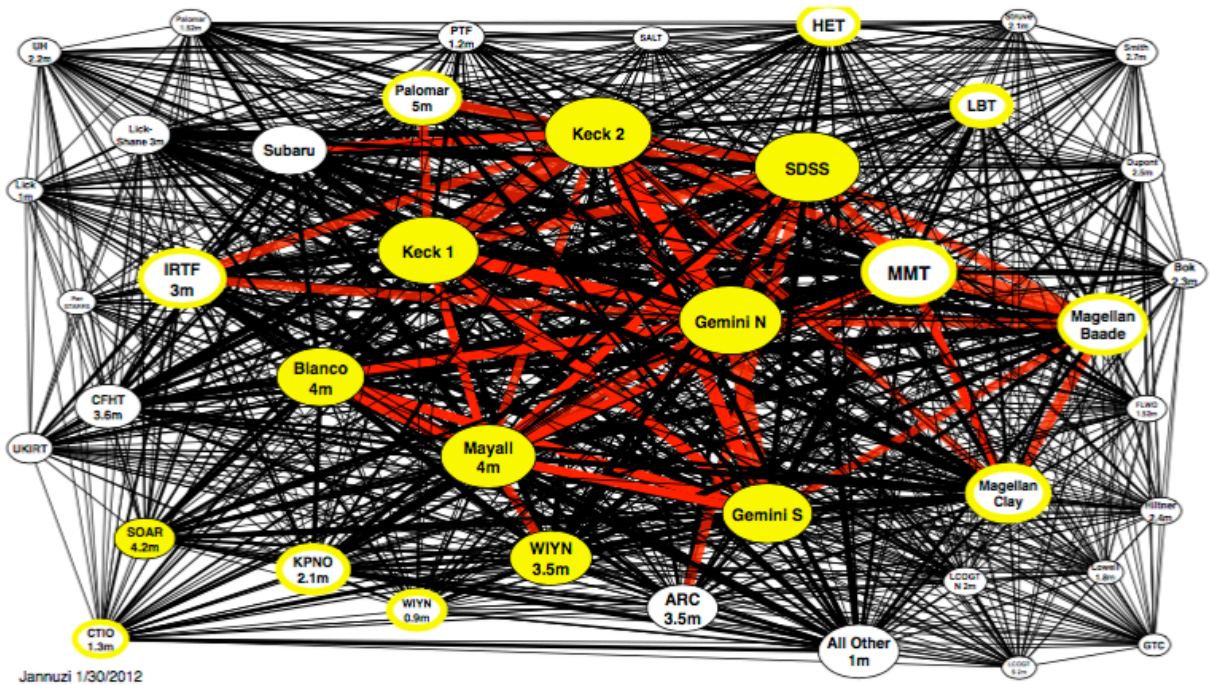

Figure 3. The connectivity of OIR telescopes in the US system. The figure results from data collected by the ground-based System Roadmap Committee's survey taken in November 2011. The US telescopes shown were used by more than 3% of the approximately 1000 US-based survey respondents. The size of the ellipse representing telescopes is based on the number of users and the thickness of the connecting lines on the number of common users. Figure from Jannuzi [24].

### 6.1 Lessons learned

Outreach and community dialogue drove the ALTAIR process. While ALTAIR was a powerful statement by the US community, the impact on the IGO is not as easily tracked. Several of ALTAIR's key recommendations remain incomplete. A major recommendation was to set up committees designed to provide communication between Gemini and the US community. We discuss these below. Another recommendation was to change Gemini governance so that Gemini was more responsive to the US community. A lesson from the earliest days of the Gemini program is that the US NGO is part of NOAO and not part of Gemini. The US NGO connection to the Gemini Board is indirect, so the US NGO by itself is a poor messenger. It is not clear how communications will be improved by the reorganization of NOAO and Gemini into one AURA management organization, NCOA. Communication between NOAO and Gemini staff has been and remains good and should improve under NCOA. However, the IGO ultimately answers to the Gemini Board.

In the post-ALTAIR period, a Gemini Caucus was organized. The group meet "via telecon to discuss and comment on the US perspective on matters that bear on the scientific quality and productivity of the Gemini telescopes"

(http://ast.noao.edu/about/committees/us-gemini-caucus). The Gemini Caucus disappeared about a year after being created. Lessons learned from the Gemini Caucus need to be gathered. If the Gemini Caucus was productive, a similar communications channel should be opened between the US board members and the US community.

The US Ground-based Optical/Near-IR System Roadmap Committee was a standing advisory committee to NOAO charged to "annually assess the state of all ground-based optical/IR telescopes operated by US institutions, including both federal and non-federal facilities and to make recommendations regarding capabilities needed by the community on near and long term timescales." This committee was a successor to the ALTAIR committee. The objective was to help the community and funding agencies maximize the scientific return of the whole system. The head of the US NGO was an ex-officio member of this committee. Again, this committee has disappeared, and lessons learned need to be collected. This kind of system survey would be especially useful as LSST reaches operational stage.

In his 1998 study of the queue, Boroson [10] discussed how the queue was productive but had failings with user satisfaction. Twenty years later these lessons have not been fully addressed. The queue is run by Gemini so is out of the control of the US NGO. However, user feedback is a NGO issue. We suggest that queue satisfaction be made a topic for NGO meetings.

In the early 2000s, US delegates to various Gemini committees regularly met with the head of the US NGO. We do not know why this stopped. With e-meetings now a common method of communication, it seems a good time to reinstate regular communication between the UG NGO and US members of Gemini committees. Similarly, a Gemini Data Reduction Group was founded in 2006 but it has not met since 2009. A status report on data reduction seems overdue.

### 6.2 Goals

The current staffing level of the US NGO is small. Required functional activities take up most of the effort, leaving little time for additional projects. One of the possible outcomes of the merger of NOAO and Gemini management into NCOA could be the disappearance of the US NGO. We believe this would be counterproductive for US astronomy. The US community needs US representation on the Gemini TAC (ITAC) and the OpsWG. We also feel that one of the most important of the seven original goals of the US NGO was outreach and community dialogue. A dialogue goes in two directions, and information needs to go to Gemini as well as to the Gemini users. The US NGO, under separate governance from Gemini, is uniquely well placed to be an honest broker. To this end we have a list of recommendations.

Personal connections to community members should be fostered. This includes continued response to HelpDesk (albeit with modern software), technical responses, and pre-observing questions. The US NGO should re-engage in doing Phase IIs. This would foster interaction with the users as well as with Gemini staff engaged in the same task.

The US NGO should feature itself as advocates for the US Gemini community. The US NGO should regularly survey each US user of Gemini to get their comments on the quality of the data they received and to listen to their suggestions for improving the output of Gemini. The visitor instrument program at Gemini has been very successful in bringing innovative capabilities to the telescopes. The US NGO should advertise this capability.

The US NGO is a division of the NOAO, soon to be NCOA, Community Science and Data Center (CSDC) division. The CSDC has an oversight committee. At least one member of this committee should be someone familiar with Gemini who can work with the US NGO. Similarly, the Gemini Observatory has a User Committee. The US NGO should have at minimum a non-voting representative on this committee to facilitate communication.

An NRC committee led by Elmegreen et al. [22] on the US OIR system in the era of LSST recommended that "the National Science Foundation should work with its partners in Gemini to ensure that Gemini South is well positioned for faint-object spectroscopy early in the era of Large Synoptic Survey Telescope operations." Elmegreen et al. also recommended that "the National Science Foundation direct its managing organizations to enhance coordination among the federal components of medium- to large-aperture telescopes in the Southern Hemisphere, including Gemini South, Blanco, the Southern Astrophysical Research (SOAR) telescope, and the Large Synoptic Survey Telescope (LSST), to optimize LSST follow-up for a range of studies." The US NGO should follow this process and advocate for the required facilities.

Workshops on topics of future interest to the US community and Gemini should be undertaken. The Kavli report [25] noted a critical need for "a community-wide planning process to motivate and review the development of the ground-based OIR System capabilities that will be needed to maximize LSST science … [including] infrastructure and computing." The US NGO is in a special position to facilitate this with connections to the Gemini staff as well as the community. Recent meetings of this kind that involved US NGO staff were "Spectroscopy in the Era of LSST" in 2013 and the Kavli report. The NCOA management model should foster this kind of activity.

As LSST approaches operation, Gemini South is increasingly a focal point for planning for LSST follow-up. However, both Gemini telescopes are vital parts of the US system of telescopes (Figure 3). The US NGO should work diligently to advocate for both of these facilities. The Gemini North site arguably provides the best ground-based seeing and infrared conditions on the planet. Gemini was designed to take advantage of this, but very few observations are now done at Gemini longward of 2.5 microns. The US NGO should investigate this issue and possibly organize a meeting of the US infrared community to discuss ground-based infrared astronomy. A broader issue is what classes of observations can be best or perhaps uniquely carried out using Gemini North.

## ACKNOWLEDGEMENTS

We acknowledge our use of the NOAO Annual Reports as a rich source of research material that was drawn upon for this report.